# On the capacitive properties of individual microtubules and their meshworks

Aarat Kalra[a], Sahil Patel[b], Asadullah Bhuiyan[b], Jordane Preto[a], Kyle Scheuer[b], Usman Mohammed[c], John Lewis[d], Vahid Rezania[c], Karthik Shankar[b,*], Jack A. Tuszynski[a,d]

[a] Department of Physics, University of Alberta, 11335 Saskatchewan Dr NW, Edmonton, Alberta T6G 2M9, Canada

[b] Department of Electrical and Computer Engineering, University of Alberta, 9107–116 St, Edmonton, Alberta T6G 2V4, Canada

[c] Department of Physical Sciences, MacEwan University, Edmonton, Alberta, T5J 4S2, Canada.

[d] Department of Oncology, University of Alberta, Edmonton, Alberta, T6G 1Z2, Canada.

[*]Corresponding author: Karthik Shankar, kshankar@ualberta.ca

**ABSTRACT**

Microtubules are hollow cylindrical polymers composed of the highly negatively-charged (~23e), high dipole moment (1750 D) protein α, β- tubulin. While the roles of microtubules in chromosomal segregation, macromolecular transport and cell migration are relatively well-understood, studies on the electrical properties of microtubules have only recently gained strong interest. Here, we show that while microtubules at physiological concentrations increase solution capacitance, free tubulin has no appreciable effect. For a particular microtubule concentration, we were able to quantify these effects by determining the capacitance and resistance of a single 20 μm-long microtubule to be $1.86 \times 10^{-12}$ F and $1.07 \times 10^{12}$ Ω, respectively. Further, we observed a decrease in electrical resistance of solution, with charge transport peaking between 20-60 Hz in the presence of microtubules, consistent with recent findings that microtubules exhibit electric oscillations at such low frequencies. Our results show that in addition to macromolecular transport, microtubules also act as charge-storage devices through counterionic condensation across a broad frequency spectrum. We conclude with a hypothesis of an electrically-tunable cytoskeleton where the dielectric properties of tubulin are polymerization-state dependent.

# INTRODUCTION

Microtubules (MTs) are cylindrical polymers composed of the heterodimers of protein α, β- tubulin that play a variety of well-recognised intracellular roles, such as maintaining the shape and rigidity of the cell, aiding in positioning and stabilisation of the mitotic spindle for allowing chromosomal segregation, acting as 'rails' for macromolecular transport and forming cilia and flagella for cell movement. Since the tubulin dimer possesses a high negative electric charge of ~23e and a large intrinsic high dipole moment of 1750 D [1-2], MTs have been implicated in electrically-mediated biological roles [3-6]. They have been modelled as nanowires capable of enhancing ionic transport [7-8], and simulated to receive and attenuate electrical oscillations [4, 9-11]. In solution, MTs have been shown to align with applied electric fields [2, 12-16]. Recently, MTs have also been modelled as the primary cellular targets for low-intensity (1-2 V), intermediate-frequency (100-300 kHz) electric fields that inhibit cancer cell proliferation, in particular glioma [17-19]. Indeed, MTs have been reported to decrease buffer solution resistance [12-13], leading to a conductance peak at TTField-like frequencies [20]. While these studies show that MTs are highly sensitive to external electric fields, answers to the questions 'How do MTs effect a solution's capacitance?' and 'What is the capacitance of a single MT?' are still elusive and crucial to the determination of the dielectric properties of living cells. The tubulin concentration in mammalian cells varies in the micromolar range (~10-25 μM) [21-22]. *In vitro*, polymerizing tubulin at such high concentrations can lead to the formation of entangled meshworks, confounding quantification of the individual MT response to electric fields. Electro-rotation, dielectrophoresis and impedance spectroscopy are thus performed using low concentrations of tubulin, in the nanomolar regime, to enable robust observation of individual MTs.

MT formation and stability are known to be optimal in buffers with ionic strength between 80 and 100 mM [23-24]. A background of BRB80 (which consists of 80 mM PIPES, 2 mM $MgCl_2$ and 0.5 mM EGTA), is thus used to study the dynamics and mechanical properties of MTs. To study their electrical properties however, the usage of such high ionic-strength solutions has historically been problematic because any dielectric attenuation caused by MTs is overwhelmed by the noise and high conductivity from the background. In the low-frequency regime (1 Hz-100 kHz), two approaches have thus far been used to estimate the dielectric properties of MTs and tubulin. One is to electrically observe low concentrations of MTs (tubulin concentration in the nanomolar regime) in the presence of low ionic strengths [12-13, 20, 25-26]. Such studies overlook the intrinsic ionic concentration of mammalian cytosol, which varies between 200 to 500 mM depending on the cell type [27-28]. Another approach to electrically interrogate MTs is to dry them: the conductivity of the buffer is nullified by evaporation, leaving polymeric tubulin behind [29-30] In a physiological situation however, MTs are solvated by the highly conductive and noisy cytosol.

In this paper, we report on our efforts overcome the barrier posed by a high ionic strength by performing electrochemical impedance spectroscopy (EIS) on cellular concentrations of tubulin. We have been able to successfully observe differences in impedance using a background of BRB80, which contains within an order of magnitude of physiologically measured ionic concentrations. Surprisingly, we find that MTs increase the solution capacitance of BRB80 whereas free tubulin does not, implicating a difference in electrical properties based only on the morphology of this protein solute. We also report a 'reversal' in the resistive behaviour of MTs compared to BRB80, with a reduction in solution resistance peaking at 20-60 Hz, a finding consistent with recent reports showing that polymerized tubulin quasi-resonantly responds to electric oscillations at ~39 Hz [9-10]. Using an equivalent circuit model for MTs, we experimentally determine the capacitance and resistance of the MT meshwork to be $1.27 \times 10^{-5}$ F and $9.74 \times 10^4$ Ω at physiological concentrations of tubulin. Using approximations, we quantify the capacitance and resistance for an individual MT as $1.86 \times 10^{-12}$ F and $1.07 \times 10^{12}$ Ω. Our values are in agreement with previous calculations and indicate that the polymerization of tubulin into MTs alters spatial and temporal charge distribution, altering the electrical properties through charge storage in the cell.

**RESULTS**

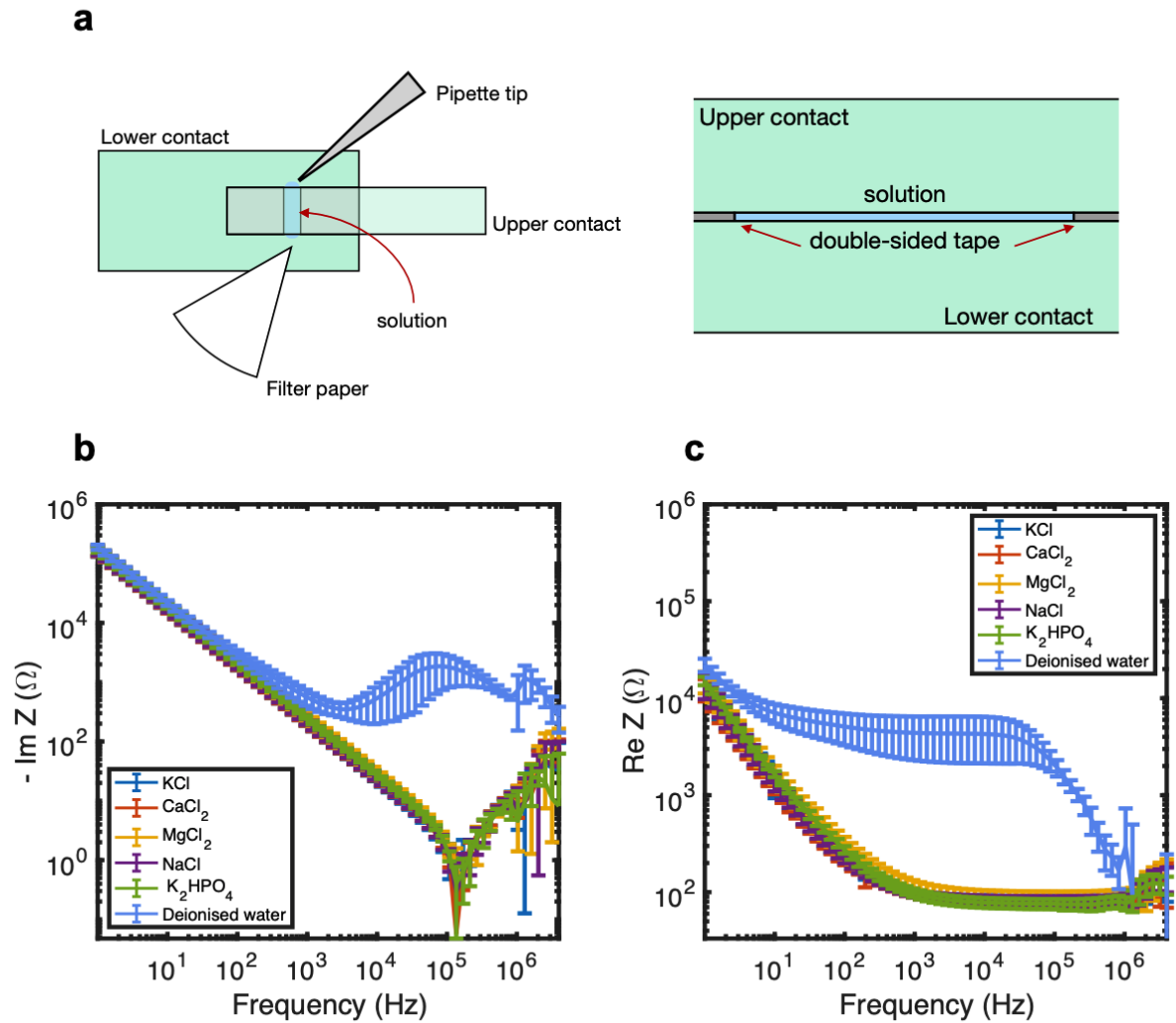

Figure 1. A parallel-plate contact device to measure the impedance properties of MTs compared to tubulin. The operation of the parallel plate device showing (a) top view (left) and side view (right). The upper and lower contacts, double-sided tape and solution are labelled in green, gray and blue, respectively. (b) Imaginary component of impedance for electrolytic solutions at 100 mM and de-ionized water. (c) Real component of impedance for electrolytic solutions at 100 mM and de-ionized water. Data display average values collected between 15 and 21 times. Error bars represent standard deviation. For measurement setup and perfusion protocols, see Materials and Methods.

**A parallel-plate contact device can accurately measure dielectric properties of ionic solutions such as the cytosol**

To determine the differences in the dielectric properties of solution caused by the presence of MTs, we aimed to create an electrode geometry that would be experimentally robust and easily modelled. We fabricated an FTO-coated parallel-plate contact device (Figure. 1a, Materials and Methods), which allowed EIS using a solution-exchange method.

We started by performing EIS on electrolytes found in the cytosol and observed a decrease in the imaginary component of impedance as a function of decreasing input frequency (Figure. 1b). The total impedance of our system was given by:

$$Z = r_c + \frac{r_s}{1+(r_s\omega C)^2} + j\left(\omega L_c - \frac{r_s^2 \omega C}{1+(r_s\omega C)^2}\right) \quad (1)$$

Here, Z is the impedance, ω is the angular frequency (given by $2\pi f$ where $f$ is the input voltage frequency), $C$ is the system capacitance, $L_c$ is the cable inductance, $r_s$ and $r_c$ are the solution and cable resistances respectively. We also observed a decrease in the real component of impedance as a function of decreasing input frequency (Figure 1c). Such a trend is expected from Warburg impedance [32-33] and follows the equation:

$$Z_{complex} = \frac{A_\omega}{\sqrt{\omega}} + \frac{A_\omega}{j\sqrt{\omega}} \qquad (2)$$

where $Z_{complex}$ is complex impedance and $A_\omega$ is the Warburg coefficient. Our circuit simplifies to the equation below if we ignore the effect of cable inductance $\omega L_c$, at frequencies below $10^5$ Hz:

$$Z = r_c - j\frac{1}{\omega C} \qquad (3)$$

Our results emulated previous data [34-36] and validated the experimental setup for further analysis.

**3.2 MTs increase solution capacitance compared with background, while unpolymerized tubulin does not have a significant effect**

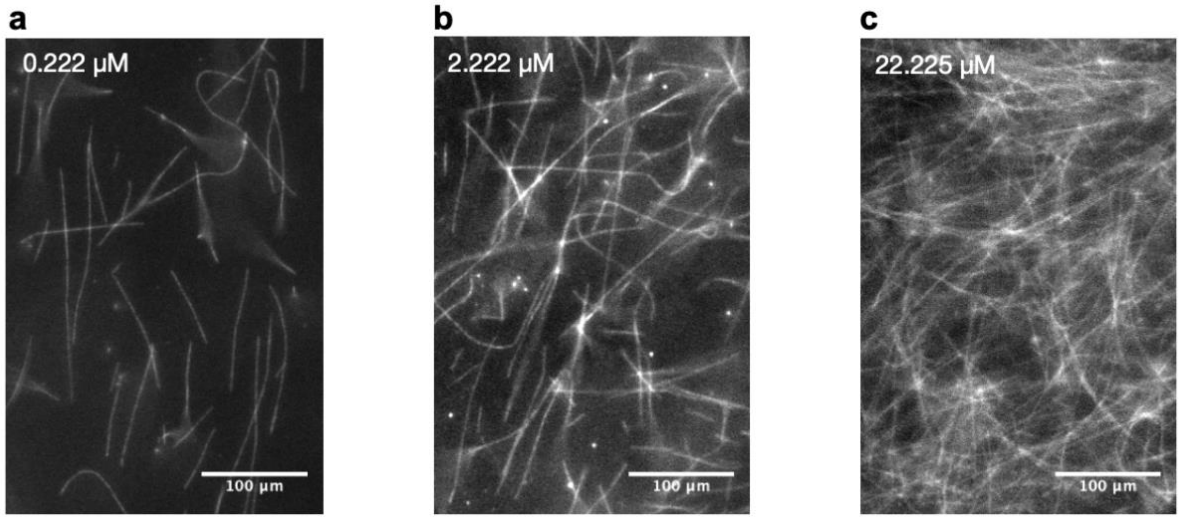

Figure 2. Microtubule imaging at different tubulin concentrations. Polymerization was performed using 45 µM tubulin, and MTs were stabilized with 50 µM paclitaxel, and subsequently diluted to a final concentration of (a) 0.222 µM tubulin (b) 2.222 µM tubulin (c) 22.225 µM tubulin, respectively. For imaging setup and polymerization protocols, see Materials and Methods.

We reconstituted and polymerized fluorescent tubulin from a stock of 45.45 μM tubulin solution (Materials and Methods). MTs were stabilised using 50 μM paclitaxel [37-38] and imaged using an epi-fluorescence microscope. On diluting MT concentration across three orders of magnitude (0.222, 2.222 and 22.225 μM tubulin), we observed that while individual MTs at low concentrations were separated by large distances, those at cellular concentrations formed enmeshed networks reported previously (Figure 2 a, b, c) [39]. Such interconnected MT meshworks are utilized by molecular motors for long-range macromolecular transport [40-41]. Here, their presence demonstrated successful MT polymerization for electrical characterization.

We performed EIS on BRB80, BRB80T (BRB80 supplemented with 50 μM paclitaxel; background for all MT-containing solutions), and MT-containing solutions in increasing order of concentration (Figure S3 a, b). We subtracted impedance values obtained for BRB80T alone from those in MT containing solutions, to determine the MT contribution to impedance. Our results showed that with an increasing MT concentration, the value of imaginary impedance became more negative, resulting in positive impedance differences (Figure 3a, b, c, d). This effect was highest at 1 Hz and decreased with a decreasing input frequency (Figure S4 a, b, c, d, e, f). Experiments with unpolymerized tubulin at the same concentrations were performed using the identical procedure, but using BRB80C (BRB80 was supplemented with 50 μM colchicine) as a background, to prevent MT nucleation [42-43]. Results with tubulin did not show an appreciable deviation at any concentration. Based on the above we can conclude that polymerization of tubulin into MTs alters their ensemble electrical properties, increasing the solution's capacitance on forming MTs and their meshworks. An increase in the solution's capacitance due to MTs has previously been modelled, [7, 44-45] indicating an increase in charge storage as free tubulin polymerizes.

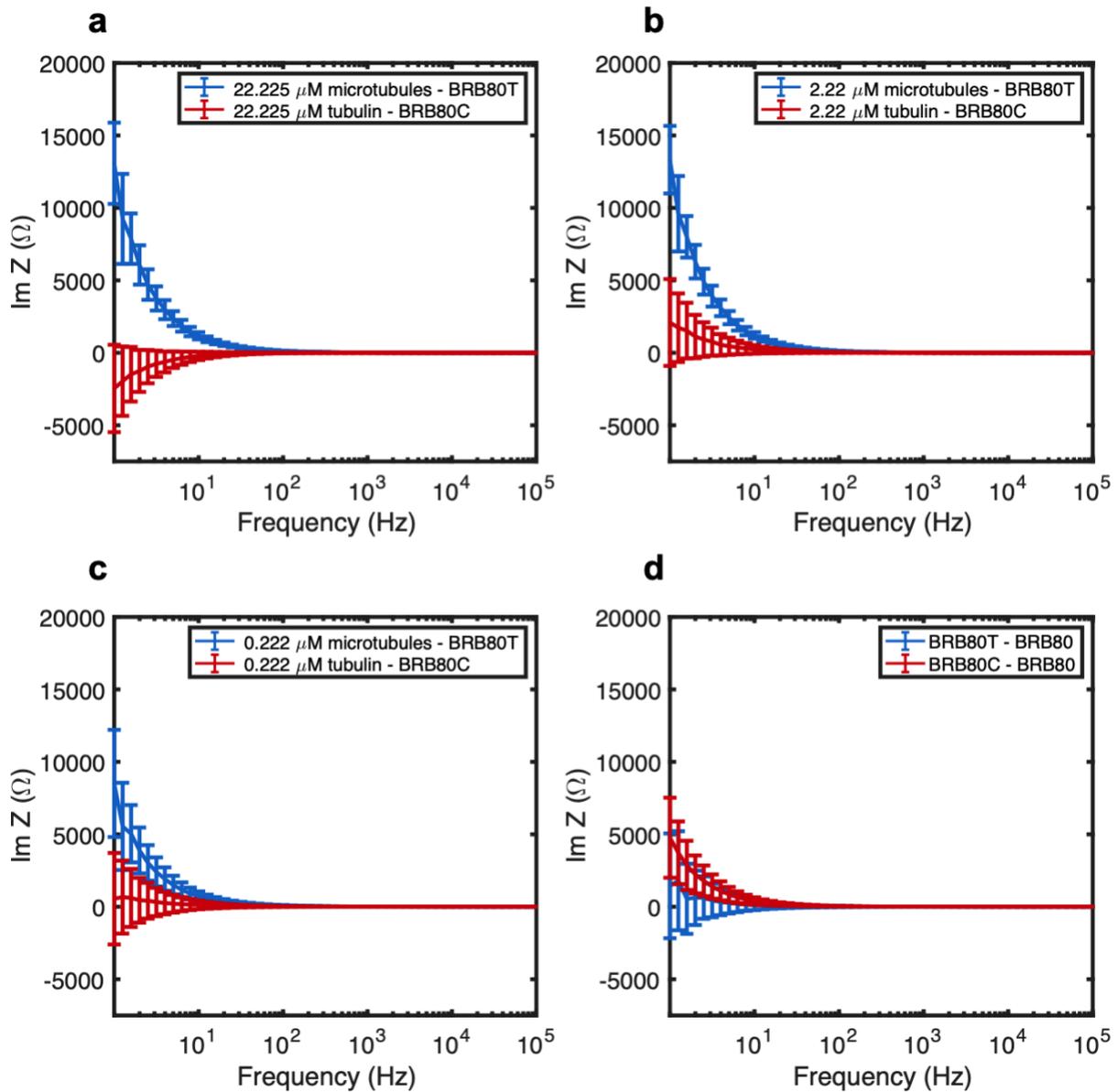

Figure 3. Mean differences in the imaginary component of impedance as a function of decreasing input AC frequency at total tubulin concentrations of (a) 22.225 µM (n = 22 experiments for tubulin, n = 21 for MTs), (b) 2.222 µM (n = 35 experiments for tubulin, n = 49 for MTs) (c) 0.222 µM (n = 35 experiments for tubulin, n = 49 for MTs), (d) comparison of the effects of paclitaxel

(BRB80T) and colchicine (BRB80C, n = 49 experiments for BRB80T, n = 35 for BRB80C, n = 84 experiments for BRB80). Error-bars represent standard deviation.

**Microtubules increase solution resistance compared to background with this effect's reversal observed at low frequencies**

Next, we looked for differences between MTs and tubulin in the real component of impedance (solution resistance). Previous studies using nanomolar tubulin concentrations and low ionic strengths (1-12 mM) have indicated that MTs enhance charge-transport in solutions [13, 20, 46]. To evaluate if this observation held true at physiologically relevant tubulin concentrations and at higher ionic strengths, we also analyzed the real component of impedance. Addition of both MTs and tubulin generally led to solution led to an increase in resistance (Figure S7 a, b, c, d), with MTs having a higher resistance at low frequencies (1-20 Hz) compared to unpolymerized tubulin. Unexpectedly, a 'reversal' of this behaviour was observed at higher frequencies as MTs began to lower solution resistance compared to tubulin (Figure S5 a, b, c, d, e, f). The reversal took place gradually between 10 and 300 Hz, with a peak between 20 and 60 Hz (Figure 4 a, b, c). Interestingly, within this range, we also found that the addition of MTs lowered solution resistance compared to background buffer BRB80T.

A reversal frequency in resistance between MTs and tubulin has not been reported before and displays the utility of our 'cell-like' approach, because the extent of this reversal decreased with decreasing concentration and was not displayed in the imaginary component of impedance or when BRB80C and BRB80T were compared. It is worth noting that recent findings using this approach showed that MT bundles and tubulin sheets generate electrical signals at ~39 Hz [9-10], and predicted an increase in solution conductance in the range that we observe, indicating that the conductance behaviour at such frequencies was due to MT- generated electrical oscillations.

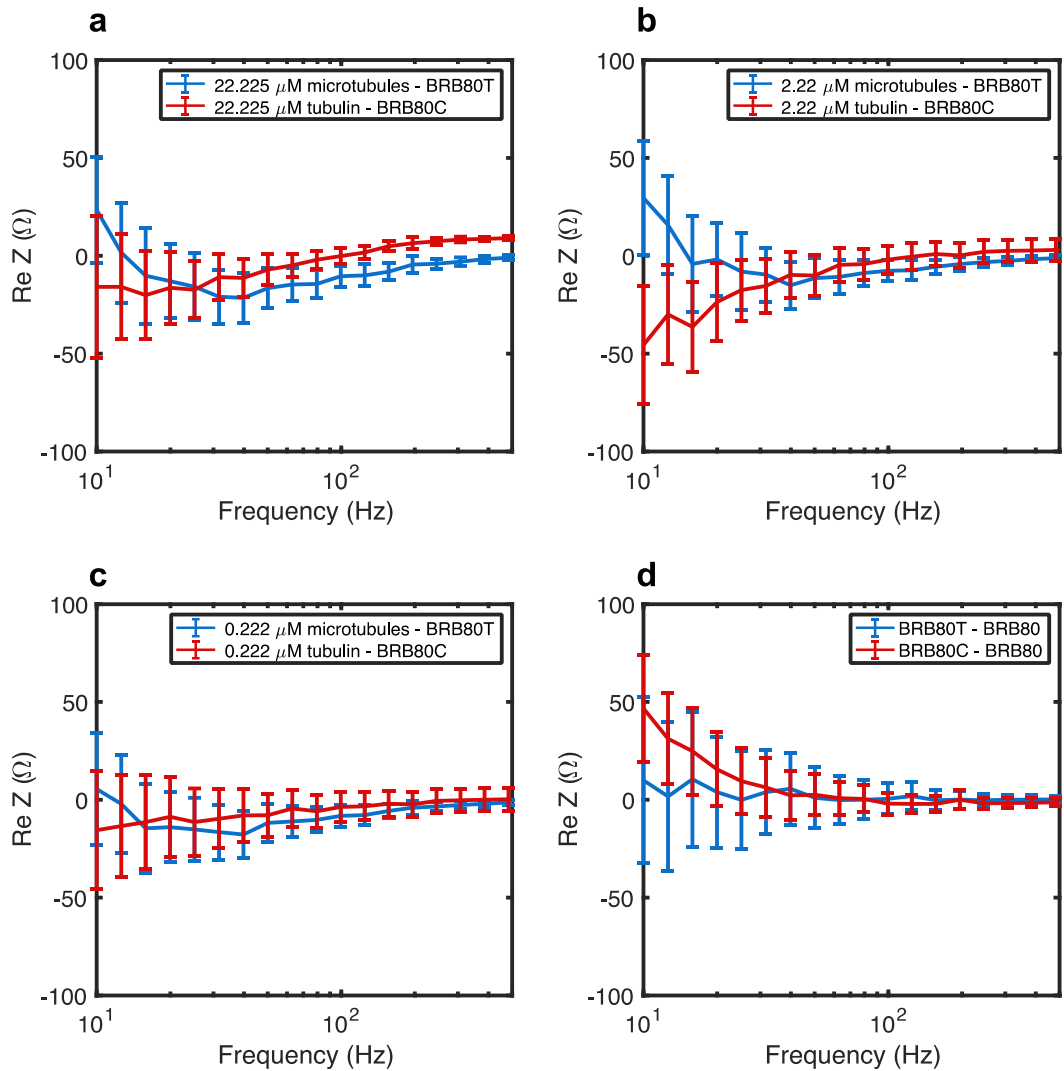

Figure 4. Mean differences in the real component of impedance as a function of decreasing input AC frequency at total tubulin concentrations of (a) 22.225 µM, (b) 2.222 µM, (c) 0.222 µM, (d) comparison of the effect of paclitaxel and colchicine on impedance. Error-bars represent standard deviation.

**The microtubule meshwork can be described as an RC circuit in parallel**

Using our experiments, our next aim was to quantify the electrical properties of a single MT. Impedance difference curves appeared linear on log-log plots with a slope of approximately negative unity, suggesting that the MT meshwork resulted in the addition of a capacitive element to the solution. We examined several combinations but a parallel RC circuit to represent the entire MT meshwork provided that best fit to observed curves (Figure 5):

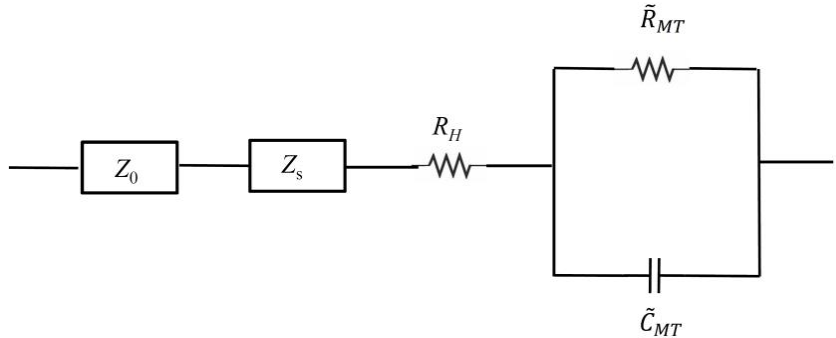

Figure 5. The equivalent electrical circuit model representing the microtubule meshwork as a parallel RC circuit, with meshwork resistance $\tilde{R}_{MT}$ and capacitance $\tilde{C}_{MT}$. The external element has impedance $Z_0$, while solution has impedance $Z_s$. $R_H$ is the small constant resistance that is ascribed to small fraction of unpolymerized tubulin that is present in MT containing solutions.

We modelled the impedance caused by external circuit elements and BRB80T as $Z_0$ and $Z_s$ respectively, as shown in Fig 5. The net impedance of the background BRB80T was thus given by:

$$Z_{\text{buffer}} = Z_0 + Z_s \tag{4a}$$

Denoting the impedance, resistance and capacitance of the entire MT meshwork by $\tilde{Z}_{\text{MT}}$, $\tilde{R}_{MT}$ and $\tilde{C}_{MT}$ respectively, the impedance for the circuit with MTs is given by:

$$Z_{\text{MT+buffer}} = Z_0 + Z_{\text{s}} + R_{\text{H}} + \tilde{Z}_{\text{MT}} \qquad (4b)$$

where,

$$\frac{1}{\tilde{Z}_{\text{MT}}} = \frac{1}{\tilde{R}_{\text{MT}}} + j\omega \tilde{C}_{\text{MT}}$$

Additionally, the impedance differences between solutions with and without MTs are given by:

$$\Delta Z = Z_{\text{MT+buffer}} - Z_{\text{buffer}} = R_{\text{H}} + \tilde{Z}_{\text{MT}} \qquad (5)$$

where

$$\tilde{Z}_{\text{MT}} = \frac{\tilde{R}_{\text{MT}}}{1 + (\omega \tilde{C}_{\text{MT}} \tilde{R}_{\text{MT}})^2} - j \frac{\omega \tilde{C}_{\text{MT}} \tilde{R}_{\text{MT}}^2}{1 + (\omega \tilde{C}_{\text{MT}} \tilde{R}_{\text{MT}})^2} \qquad (6)$$

We subsequently fit experimental impedance difference curves shown in Figure 3 and Figure S7 to real and absolute value of imaginary parts of $\Delta Z$ using $R_{\text{H}}$, $\tilde{R}_{MT}$ and $\tilde{C}_{MT}$ as our fit parameters. Here, $R_{\text{H}}$ is a resistance ascribed to the nominal fraction of unpolymerized tubulin present in MT containing solutions. The fitted curves are displayed in Figure 6 and the optimal fit parameters are listed in Table 1 (see Materials and Methods for details).

| Tubulin] (μM) | $\tilde{C}_{MT}$ (F) | $\delta \tilde{C}_{MT}$ (F) | $\tilde{R}_{MT}$ (Ω) | $\delta \tilde{R}_{MT}$ (Ω) | $\tilde{R}_H$ (Ω) | $\delta \tilde{R}_H$ (Ω) |
|---|---|---|---|---|---|---|
| 22.222 | $1.27 \times 10^{-5}$ | $1.48 \times 10^{-7}$ | $9.74 \times 10^{4}$ | $1.18 \times 10^{4}$ | 2.12 | 40.61 |
| 2.222 | $1.25 \times 10^{-5}$ | $1.67 \times 10^{-7}$ | $1.00 \times 10^{5}$ | $1.40 \times 10^{4}$ | 0.61 | 34.79 |
| 0.222 | $2.01 \times 10^{-5}$ | $3.38 \times 10^{-7}$ | $9.97 \times 10^{4}$ | $2.82 \times 10^{4}$ | 0.41 | 31.95 |

Table 1. Fit parameters attained by fitting the real and imaginary components of impedance to equation 6. Fit parameters represent effective capacitance $\tilde{C}_{MT}$, and resistance $\tilde{R}_{MT}$ introduced into the solution through the addition of the MT meshwork at different concentrations. $R_H$ is the small constant resistance that is ascribed to small fraction of unpolymerized tubulin that is present in MT containing solutions. $\delta \tilde{R}_{MT}$, $\delta \tilde{C}_{MT}$ and $\delta \tilde{R}_H$ correspond to 95% confidence intervals for the fit parameters. Corresponding graphs are displayed in Figure 6.

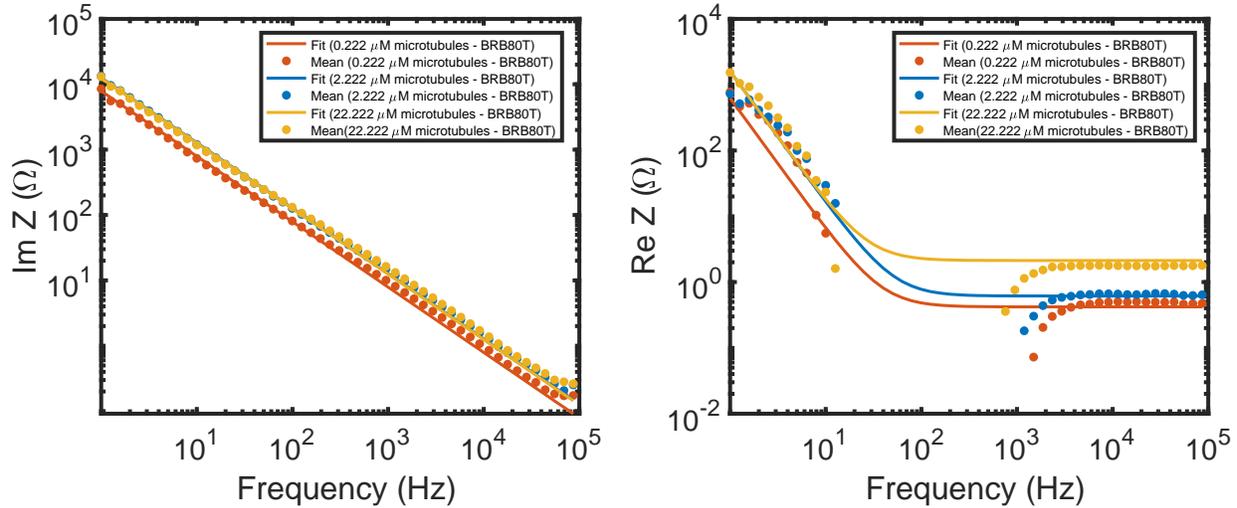

Figure 6. Mean differences of (a) imaginary and (b) real impedance curves for 0.222 μM, 2.222 μM and 22.222 μM, are fitted with the model described in Eq. 5 and Figure 5. Fit parameters and confidence intervals are displayed in Table 1. For detailed fitting methodology, see Materials and Methods.

**Experimentally derived values for the capacitance and resistance of a single microtubule match theoretical predictions.**

To determine the capacitance and resistance of a single MT from corresponding meshwork mean values, we made two assumptions. (1) The average length of an MT was taken as 20 μm based on visual inspection and (2), MTs were oriented in parallel to one another at 0.222 μM concentration (Figure 2 a). We did not extend the second assumption to concentrations of 2.22 μM and 22.22 μM due to the complex nature of the meshwork, for which a superficial series and/or parallel circuit combination would be inadequate. However, our simple assumptions do provide an upper limit for the capacitance of a single MT. Given the presence of $N = 1.08 \times 10^7$ MTs in our solution volume at the 0.222 μM concentration, the capacitance for a single 20 μm long MT was calculated to be $C_{MT} = \tilde{C}_{MT}/N = 1.86 \times 10^{-12}$ F. We extended our set of assumptions to calculate the resistance of a single 20 μm long MT at 0.222 μM concentration. Here, the total resistance of the MT meshwork would be given by $\tilde{R}_{MT} = R_{MT}/N$ where each MT has resistance $R_{MT}$, implicating $R_{MT} = \tilde{R}_{MT} \times N = 1.07 \times 10^{12} \Omega$.

To determine if these values were consistent with theoretical predictions, we first calculated the Debye Length $\lambda_D$ of an MT when present in an ionic concentration $c = 80$ mM to account for an environment consisting of BRB80T.

$$\lambda_D = \frac{\lambda_D^0}{\sqrt{c}} \tag{7}$$

Here, $\lambda_D^0$ is a Debye length (calculated to be ~10 nm) in a solution with ionic concentration $c_0 = 1$ mM [13]. Based on the Debye length, the total capacitance of the MT cylinder has been calculated by previous models to be given by [8, 44]:

$$C_{MT,theor} = \frac{2\pi\varepsilon_0\varepsilon l}{\ln\left(\frac{r + \lambda_D}{r}\right)} \tag{8}$$

Substituting the value of $\lambda_D$ from equation (7) and using the Taylor series expansion for $\ln(1 + x)$, we get:

$$C_{MT,theor} = \frac{2\pi\varepsilon_0 \varepsilon l}{\ln\left(1 + \frac{\lambda_D^0}{r\sqrt{c}}\right)} \approx \frac{2\pi\varepsilon_0 \varepsilon l r\sqrt{c}}{\lambda_D^0} \qquad (9)$$

We continued our previous assumption that each microtubule has length $l = 20$ µm, radius $r = 12.5$ nm and is surrounded by an atmosphere of dielectric constant $\varepsilon = 80$. Placing these values in Equation 9, we calculated the total capacitance of an MT to be 9.8 pF. Interestingly, previous work shows that $\varepsilon < 10$ for tubulin protein [1], enabling an extrapolation of $\varepsilon = 40$ between the protein and the bulk solution [47]. The use of this intermediate value for $\varepsilon$ leads to an MT capacitance of 4.9 pF. It is worth noting that both these values are consistent with our experimentally observed value of 1.86 pF per MT.

Fluorescence images (Figure 2 a) showed that MTs orient in the plane of glass surfaces at low concentrations, indicating that resistance to ionic charge transport in solution arose from flow across the porous MT cross section, rather than along their lengths. The lower limit of Ionic current through nanopores present between adjacent tubulin dimers in two-dimensional zinc-induced sheets is previously calculated to be $0.02 fA$, at a driving voltage of 5 mV [10]. This leads to a nanopore resistance of $R_{nanopore} = 2.50 \times 10^{14}$ Ω. We note that the conditions of our study vary significantly from those in this reference: Firstly, unlike two-dimensional sheets which present ions with a single resistance, cylindrical microtubules used in our study present both MT entering and MT exiting resistances (Figure 8 a). Secondly, the study utilizing sheets was performed at an ionic concentration of 140 mM, whereas our study was performed at a corresponding concentration of 80 mM. Thirdly and importantly, the resistance calculated above applies to resonant conditions, with the driving input frequency in the range of 10-100 Hz and would reduce as the input frequency is altered.

We calculated the total resistance for ions entering and exiting transversely through an MT as $R_{MT,enter} = R_{MT,exit} = \frac{R_{nanopore}}{n} = 7.69 \times 10^9$ Ω where $n$ is the total number of nanopores (equal to the number of tubulin dimers) in a 20 µm long MT. We then addressed the first dissimilarity mentioned above, regarding MT entering and exiting impedances, by adding these resistances in series. This leads to total resistance $R_{MT} = R_{MT,enter} + R_{MT,exit} = 1.54 \times 10^{10}$ Ω for ionic movement across the cross section. We also note that theoretically, the total ionic resistance across the cross section has been calculated by [48]:

$$R_{MT,theor} = \frac{\ln\left(1 + \frac{\lambda_D}{r}\right)}{2\pi l \sigma} \approx \frac{\lambda_D}{2\pi l \sigma} \tag{10}$$

Here, the solution conductivity $\sigma$ is proportional to the ionic concentration $c$. Additionally, because $\lambda_D$ depends inversely on the square root of ionic concentration, $R_{MT,theor}$ scales with $\frac{1}{c^{3/2}}$. Thus, to correct for the second dissimilarity arising from the difference in ionic concentration, we multiplied our resistance by a factor of $\mu = \left(\frac{140}{80}\right)^{3/2} = 2.31$. Thus, our computed final value for the resistance across the MT cross section was $\mu \times R_{MT,theor} = 4.71 \times 10^{10}$ Ω. We expect the resistance of the MT to be higher in non-resonant conditions (at frequencies >100 Hz), implicating a close match to our experimentally derived value of $1.07 \times 10^{12}$ Ω. As the applied AC voltage oscillates with the frequency it switches electric field direction every half cycle, aiding the movement of in-phase ions across the MT wall, but impeding this flow in out-of-phase ions. Thus, the total resistance is also a function of the input frequency.

Ionic movement to enter the MT cylinder takes place against the potential gradient set up by the tubulin dipole in the MT wall, which could additionally account for the high resistance that we observe.

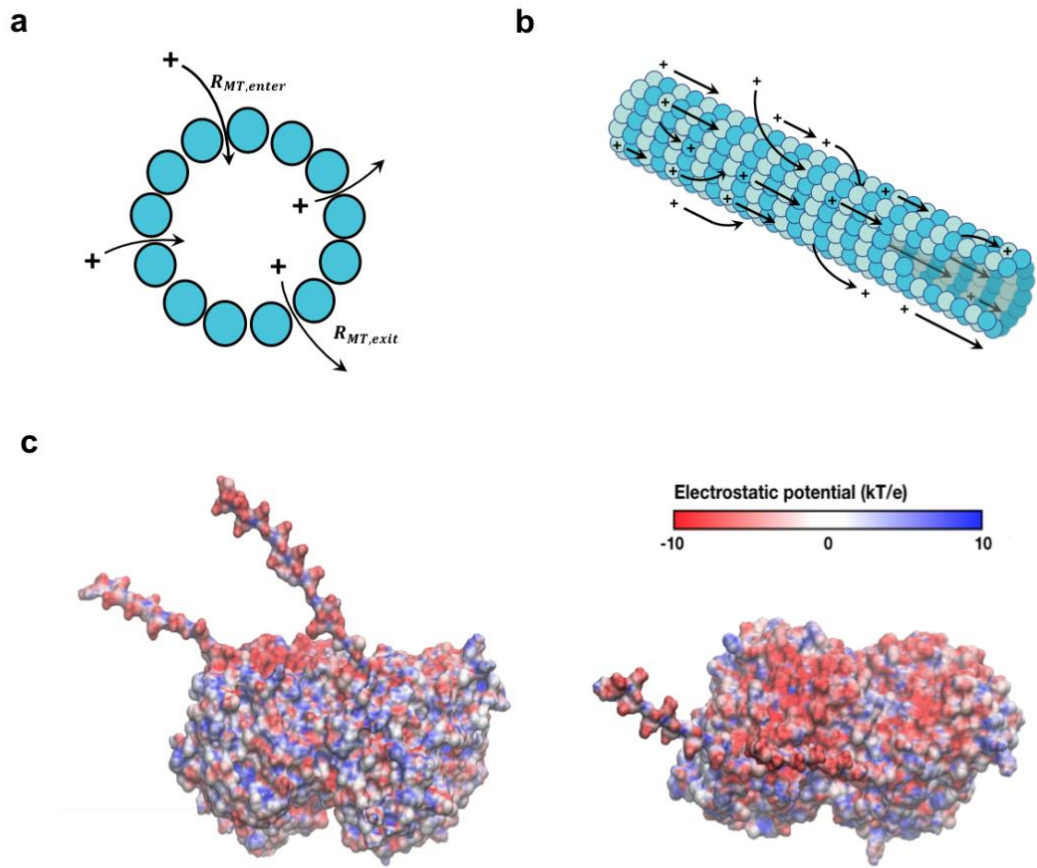

Figure 8. Schematic of charge transport along and across an MT. (a) A representation of charge flow across the MT cross section through nanopores present between adjacent protofilaments. (b) A representation of charge flow through both inner and outer modes along an MT. Arrows depict charge flow via both mechanisms, enabling MT charge storage across a broad spectrum of frequencies, and charge transport at low AC frequencies in the cell. (c) Side view (left) and top view (right) of the tubulin dimer, displaying distribution of electrostatic potential at different

locations. The negatively charged C-termini face towards the solution and contains ~50 % of the total negative charge on a tubulin dimer.

## DISCUSSION

Our measurements using a parallel plate contact device reveal interesting electrical properties of MTs at physiological concentrations. Unlike studies exposing MT-containing solutions to non-uniform electric fields [12-14, 20], our device allowed robust quantification of electrical impedance in the presence of spatially uniform electric fields. Our results show that the addition of MTs mimics a parallel RC element placed in series with the high-ionic strength solution, with a nonlinear dependence on the MT number. Unpolymerized tubulin did not alter capacitance significantly, indicating changes in electrical properties of tubulin as it polymerizes.

### The physical underpinnings of an increased capacitance

Dense counterion condensation on the MT surface has been previously extensively predicted and simulated due to a variety of reasons [7-8, 44, 49-50]. Firstly, the negative charge of the tubulin dimer attracts counterions in solution, leading to the presence of a double layer and depletion region outside the microtubule surface [7-8, 46, 50]. In addition to this, the charge distribution in the MT protein wall is highly non-uniform, with the outer surface containing approximately four times the charge compared to the inner surface [49]. This asymmetry between the inner and outer electrostatic potentials serves to enhance capacitance and is responsible for the abnormally large dipole moment of the tubulin dimer [1]. The asymmetry also manifests through C-terminal 'tails' composed of 10-12 amino-acids, that can extend 4-5 nm outwards from each tubulin monomer. These slender C-termini tails are highly negative, containing about 50 % of the charge of the tubulin dimer [51]. As they stretch outwards into the solution in a pH and ionic strength dependant manner, they increase the effective area of the tubulin dimer and significantly contribute to the overall MT capacitance [7-8].

Coherent oscillations of these C-terminal tails are modelled to generate solitonic pulses of mobile charge along the outer surface of an MT, creating ionic currents along its' length [7, 44, 52].

Ions from the bulk solution are also modelled to be pumped into the hollow MT lumen through nanopores in its' wall, resulting in charge accumulation inside the cylindrical MT over time [45]. A recent study using molecular dynamics simulations showed that the permeability of the MT lumen was significantly higher for $Na^+$ and $K^+$ as opposed to $Ca^{2+}$, allowing for free movement of selective ions into the MT lumen across its porous surface [49]. To the best of our knowledge, our findings are the first to experimentally quantify this resistance encountered by charge flow across the MT cross section. These results implicate not only ionic movement along the microtubule axis, but also across and inside it, enhancing the modelled roles of MTs as complex subcellular nanowires.

**Implications for the cell**

Our findings on cell-like complex MT networks at physiological tubulin concentrations indicate a cellular role for MTs as wires that store charge. Neuronal environments where the MT are spontaneously nucleated from free tubulin, such as growth cones, would experience large capacitance changes over short bursts of time. This ability would significantly impact action potentials, that are known to depend strongly on the local charge distributions [53]. Additionally, ionic movement across the MT wall would enhance their roles as attenuators of local cation distributions. In nonneuronal environments, transient ionic currents around on a MT during mitosis could impact MT dynamics and potentially influence chromosome segregation. Specifically, $Ca^{2+}$ ion storage/flow about an MT would trigger its' depolymerization, whereas waves of $Mg^{2+}$ or lowering in the local pH (increasing $H^+$) would lead to MT stabilization [54-55]. The attraction of $Zn^{2+}$ or $Mn^{2+}$ ions in the vicinity would lead to formation of two-dimensional tubulin polymers [56-57]. Properties of the cytoplasm such as polarizability and relative permittivity would get severely attenuated due to the presence of MTs in the vicinity. Due to the polymerization state of tubulin altering solution capacitance, our findings implicate a temporal evolution of capacitance and ionic flows as the ratio of MTs to free unpolymerized tubulin changes [58-60]. MT lattice defects, that take place when a tubulin dimer is missing in an MT wall [61-62], would cause a large ionic flux to develop at the defect site. Such localised shifts in charge distributions would be most significant at the MT end, where large localized charge fluxes would form due to sudden changes in both ionic resistance and capacitance. Free/polymerized tubulin would thus regulate local and global electrical properties, creating spatially dynamic gradients of charge storage and flux. We envision a cytoskeleton that, in addition to transporting macromolecules, stores and transports ionic signals and electrical information across the cytoplasm (Figure 8 b).

Our findings can be coupled with a vast array of bio-nanodevices that utilize MTs and MAPs (microtubule-associated proteins) for construction of bio-nanotransporters and bio-actuators (Diez 2003; Hess 2018; Li 2018; Kapitein 2018; Yokokawa 2012). In specific conditions, MAP-MT systems are capable of repositioning organelles (Kapitien 2017), directionally transporting microtubules (Yokokawa 2004; Isozaki 200) and even drive their movement within zero-mode waveguides (Yokokawa 2018) and inorganic nanotubes (Li 2014). Charge storage and transport along MTs can be coupled to mechanical MAP-based movement to develop a wide range of nano-actuators and nano-sensing devices.

**Future work**

When compared to cells, the rates of MT nucleation and polymerization are significantly lower in BRB80. This difference can be attributed to the absence of MAPs and macromolecular crowding [63-64]. Mammalian cells contain high concentrations of $K^+$ ions (140-300 mM) [27-28], which, in addition to MAPs and molecular crowding agents, would be included in a future study to attain physiological equivalence. Raw data from complex impedance will be used to derive properties of individual MTs such as the dielectric constant, refractive index, polarizability and Debye relaxation time through the Clausius-Mossotti equation. To improve the present two-contact parallel-plate device, a four-contact device will lower electrode polarization effects and enable quantification of the relative permittivity of MT and tubulin containing solutions. Our device geometry will also be used to perform DC (direct-current) measurements, determine the contribution of MTs to impedance relaxation time and evaluate the voltage dependence of capacitance on MT-containing solutions. Interestingly, this aspect has been discussed previously: the inductance of a single protofilament is calculated to be <1 fH [8]. Further investigation is required to experimentally confirm these predictions.

**CONCLUSIONS**

We used EIS to compare the complex impedance of MT- and tubulin-containing solutions. A high ionic strength buffer (BRB80) created a high noise, low impedance background, which was countered through the use of physiological concentrations of tubulin. While the presence of MTs increased solution capacitance, unpolymerized tubulin did not have any appreciable effect. We used an RC circuit to model the MT meshwork across three orders of concentrations, and found at physiological levels, the capacitance and resistance of the meshwork is to be $1.27 \times 10^{-5}$ F and $9.74 \times 10^{4}$ Ω . In a study that is the first of its' kind to the best of our knowledge, we determined the capacitance of a single MT to be $1.86 \times 10^{-12}$ F. We used our lowest, 0.222 μM concentration, for this calculation and assumed that MTs were oriented in parallel to one another. We envision a dual electrical role for MTs in the cell: that of charge storage devices across a broad frequency spectrum (acting as storage locations for ions), and of charge transporters (bionanowires) at specific, low frequencies between 20 and 60 Hz. Our findings also indicate that the electrical properties of tubulin dimers change as they polymerize, revealing the potential impact of MT nucleation and polymerization on the cellular charge distribution. Our work shows, that by storing charge and attenuating local ion distributions, microtubules play a crucial role in governing the bioelectric properties of the cell.

## MATERIALS AND METHODS

### Tubulin reconstitution

Before reconstituting tubulin solution from lyophilized protein powder, G-PEM buffer was prepared by adding 2.5 μL GTP to 247.5 μL BRB80 (Cytoskeleton Inc; BST01). 180 μL of G-PEM buffer was pipetted into lyophilized tubulin powder (Cytoskeleton Inc; T240). 20 μL of MT cushion buffer (60% glycerol in BRB80) was subsequently added. All steps were performed on ice.

Rhodamine-labelled tubulin solution was prepared by adding 4 μL of G-PEM buffer to labelled tubulin powder (Cytoskeleton Inc; TL590m) and subsequently adding 1 μL of MT cushion buffer.

A final tubulin stock solution of 45.45 µM was thus prepared, and a final labelling ratio of 1:15 was used. Finally, aliquots were snap-frozen and stored at -80 °C.

**MT polymerization and stabilization**

MT polymerization was performed by incubating 45.45 µM tubulin aliquots in a 37 °C water bath for 30 minutes. BRB80 solution was heated alongside tubulin during the first 15 minutes of polymerization. Subsequently, BRB80 was incubated at room temperature, and paclitaxel solution (Cytoskeleton Inc, TXD01; 2 mM stock) was thawed at room temperature alongside it. After 30 minutes of tubulin polymerization brought to completion, 100 µL of BRB80 was added to 5 µL of 2 mM paclitaxel. For preparing 0.222 µM MTs, 99.5 µL of this solution was added to 0.5 µL tubulin solution. For 2.222 µM MTs, 95 µL of this solution was added to 5 µL tubulin solution. For 22.225 µM MTs, 5 µL of this solution was added to 5 µL of tubulin solution. For preparing BRB80T, 45 µL of this solution was added to 45 µL of BRB80.

For tubulin stabilization, 2 µL of colchicine stock solution (Sigma-Aldrich, C9754; 5 mM in DMSO) was added to 100 µL BRB80. For preparing 0.222 µM tubulin, 99.5 µL of this solution was added to 0.5 µL unpolymerized tubulin solution. For 2.222 µM tubulin, 95 µL of this solution was added to 5 µL unpolymerized tubulin solution. For 22.225 µM tubulin, 5 µL of this solution was added to 5 µL of unpolymerized tubulin solution. For preparing BRB80C, 45 µL of this solution was added to 45 µL of BRB80.

**Fluorescence imaging of MTs**

Imaging was performed on a Zeiss Examiner.Z1 microscope using a Hamamatsu EMCCD C9100 camera, a Zeiss plan-Apochromat 1.4 NA 63x lens. After pipetting MT solution (2-5 µL) onto a glass slide (VWR 48382-173) a coverslip (VWR 48393-070) was placed on the solution, allowing it to spread. The microscope used an EXFO X-Cite 120 fluorescence source and excitation and emission filters of 535 nm and 610 nm, respectively. Exposure times between 50 ms and 300 ms were used for imaging to validate the presence of MTs.

**Electrode design and device construction**

Each 'plate' in the parallel-plate contact device was formed by FTO (Fluorine-doped Tin Oxide)-coated glass slides (Sigma Aldrich, 735140). The slides were cleaved to dimensions of 1.5 mm x 10 mm x 50 mm for the upper contact and 1.5 mm × 27 mm × 50 mm for the lower contact. The cleaving dimensions were set using 3D printed devices that were placed as holders (The Shack, University of Alberta; Figure S1). The slides were sonicated and subjected to Reactive Ion Etching (RIE) using a 5-minute exposure to oxygen plasma (Oxford Instruments, NGP80) to remove surface particulate matter. 70 μm thick double-sided tape was used as a spacer, which formed a chamber of dimensions 3 mm × 1.25 cm × 70 μm. The top electrode was placed using a separate 3D-printed holder device (Figure S1). Once the device was constructed using the above protocol, solution was perfused into the chamber using a pipette and a filter paper for suction (Video S1). We used flat copper electrode clips in a three-electrode configuration to connect to our capacitor device. The counter electrode was connected to the lower electrode, and the working and reference electrodes were connected to the top electrode of our device.

**Impedance measurements**

Experiments were conducted using Electrochemical Impedance Spectroscopy (EIS) on a Zahner Zennium impedance analyzer. The parallel-plate contact device was placed into the 3D-printed holder for stabilisation (Figure S1). The contacts from the machine were connected to the parallel-plate device using flat-faced copper alligator clips. A three-electrode configuration was used: the counter electrode was attached to the lower contact of the parallel-plate device, whereas the working electrode was attached to the upper contact with the reference electrode orthogonally clipped onto the clip of the working electrode. Within the Thales Z3.04 environment, the Potentiostat Mode was ON; the stabilisation delay was set to 1 $s$, the rest potential drift tolerance was set to 250 µ$V$, $V_{rms}$ was set to 5 mV. Solutions were perfused into the experimental chamber using a micropipette tip at one opening, and a filter paper at the other opening for suction, similar to protocols used for TIRF (Total Internal Reflection Fluorescence) microscopy[31]. The frequency range of the EIS measurement was set from 4 MHz to 1 Hz, and data were subsequently collected.

**Data analysis**

MT and tubulin samples were analyzed using data from five to seven days of experiments. Each day consisted of three to seven solutions for each concentration being tested, with one frequency sweep per solution. Readings of each sweep were saved as a .csv file, and next sample was loaded by solution exchange method. Water was run as the first solution for each day of experiments. BRB80T was run prior to MT solutions, and BRB80C were run prior to the free tubulin containing solutions. MT- and free tubulin-containing solutions were run on separate days, in increasing order of concentration. MATLAB (The Mathworks; Natick, MA) scripts were used for data analysis. Fitting to the real and imaginary components of impedance was performed using the function lsqnonlin. Initial guess values for the MT meshwork resistance and capacitance were $10^5$ F and $10^{-5}$ Ω, respectively, based on visual inspection of raw data. The initial guess values for the nominal series resistor, $R_H$, were set at 1.78, 0.6 and 0.4 Ω for with tubulin concentrations of 0.222, 2,222 and 22.222 µM respectively. The 95% confidence intervals were determined using the function nlparci. Error propagation was performed assuming no relationship between various days of data collection.

**AUTHOR INFORMATION**

**Author contributions.**

AK, JT and KS designed the paper. AK, SP, AB and UM collected the data. AK, SP, AB, KS and JP analysed and validated the data. AK, VR, KS and JT wrote the paper. AK, SP and JP made figures for the paper. KS, JT and JL provided resources for experimentation and analysis.

**Notes**

The authors declare no competing financial interests.

**Acknowledgements**

JT and KS gratefully acknowledge funding from Novocure LLC and NSERC (National Science and Engineering Research Council). AK and KS acknowledge funding from CMC (Canadian Microsystems Corporation). AK acknowledges Najia Mahdi for helping him with reactive ion etching. AB acknowledges personnel at The Shack at the University of Alberta and particularly Tristan Stark for teaching him 3D printing. AB and SP acknowledge Liam McRae for showing them how to use the Impedance Analyzer. The authors acknowledge Philip Winter, Sheng Zeng, Dr. Piyush Kar, Dr. Dayal Pyari Srivastava, Dr. Ze'ev Bomzon, Prof. Frank Hegmann and Dr. Kris Carlson for useful discussions.

Supporting Information for

# On the capacitive properties of individual microtubules and their meshworks

Aarat Kalra[a], Sahil Patel[b], Asadullah Bhuiyan[b], Kyle Scheuer[b], , Jordane Preto[a], Usman Mohammed[c] John Lewis[d], Vahid Rezania[c], Karthik Shankar[b, *], Jack A. Tuszynski[a,d]

[a] Department of Physics, University of Alberta, 11335 Saskatchewan Dr NW, Edmonton, Alberta T6G 2M9, Canada

[b] Department of Electrical and Computer Engineering, University of Alberta, 9107–116 St, Edmonton, Alberta T6G 2V4, Canada

[c] Department of Physical Sciences, MacEwan University, Edmonton, Alberta, T5J 4S2, Canada.

[d] Department of Oncology, University of Alberta, Edmonton, Alberta, T6G 1Z2, Canada.

[*]Corresponding author: Karthik Shankar kshankar@ualberta.ca

a

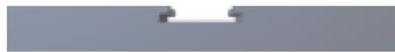 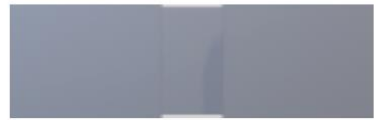

b

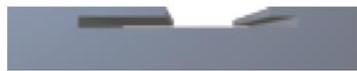 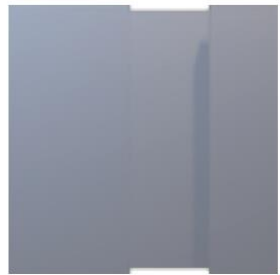

c

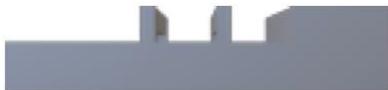 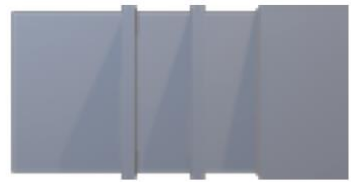

d

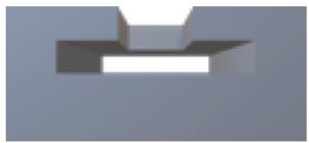 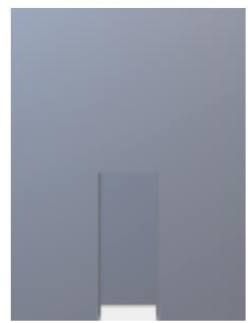

Figure S1. 3D printed holders used to fabricate and align the parallel-plate contact device. (a) Top view (left) and side view (right) of holder for the parallel plate device used to perform impedance measurements. (c) Top view (left) and side view (right) of slider used to position the double-sided tape exactly to fabricate the device. (d) Top view (left) and side view (right) of the holder used to position the upper contact precisely on the lower contact.

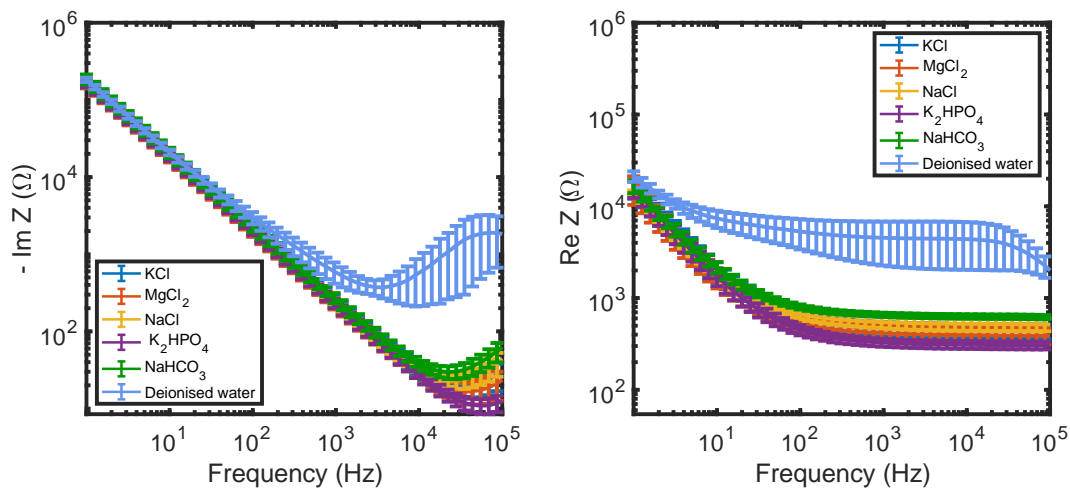

Figure S2. Validation of parallel-plate contact device using 0.5 mM electrolytic solutions (a) Imaginary component of impedance for electrolytic solutions at 0.5 mM and de-ionised water. (b) Real component of impedance for electrolytic solutions at 100 mM and de-ionised water. Data displays average values collected between 15 and 21 times. Error bars represent standard deviation.

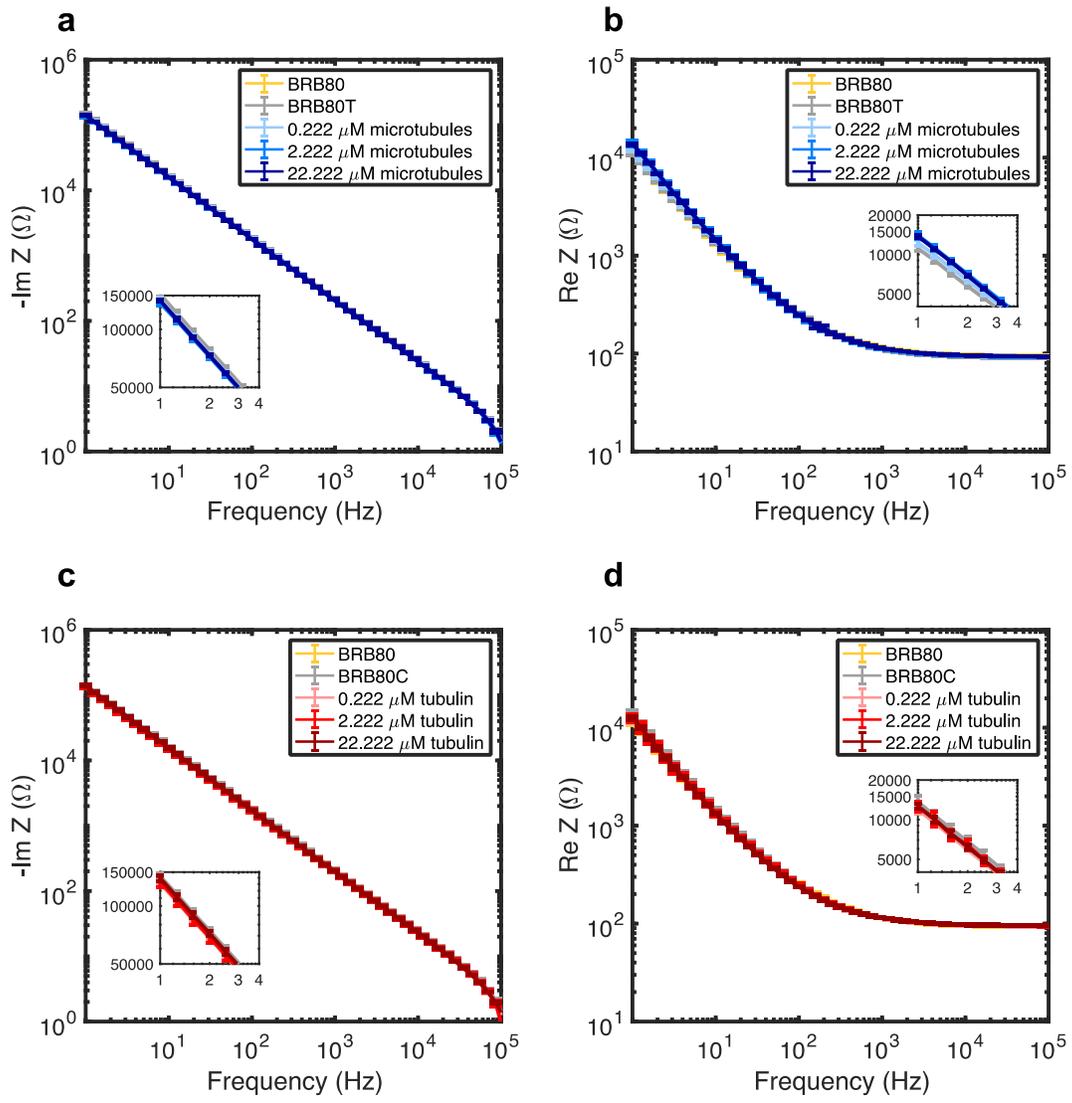

Figure S3. Example of microtubule and tubulin data for the purpose of displaying typical impedance values. Raw values for (a) imaginary and (b) real component of impedance at different

concentrations of microtubules. Raw values for (c) imaginary and (d) real component of impedance at different concentrations of unpolymerized tubulin.

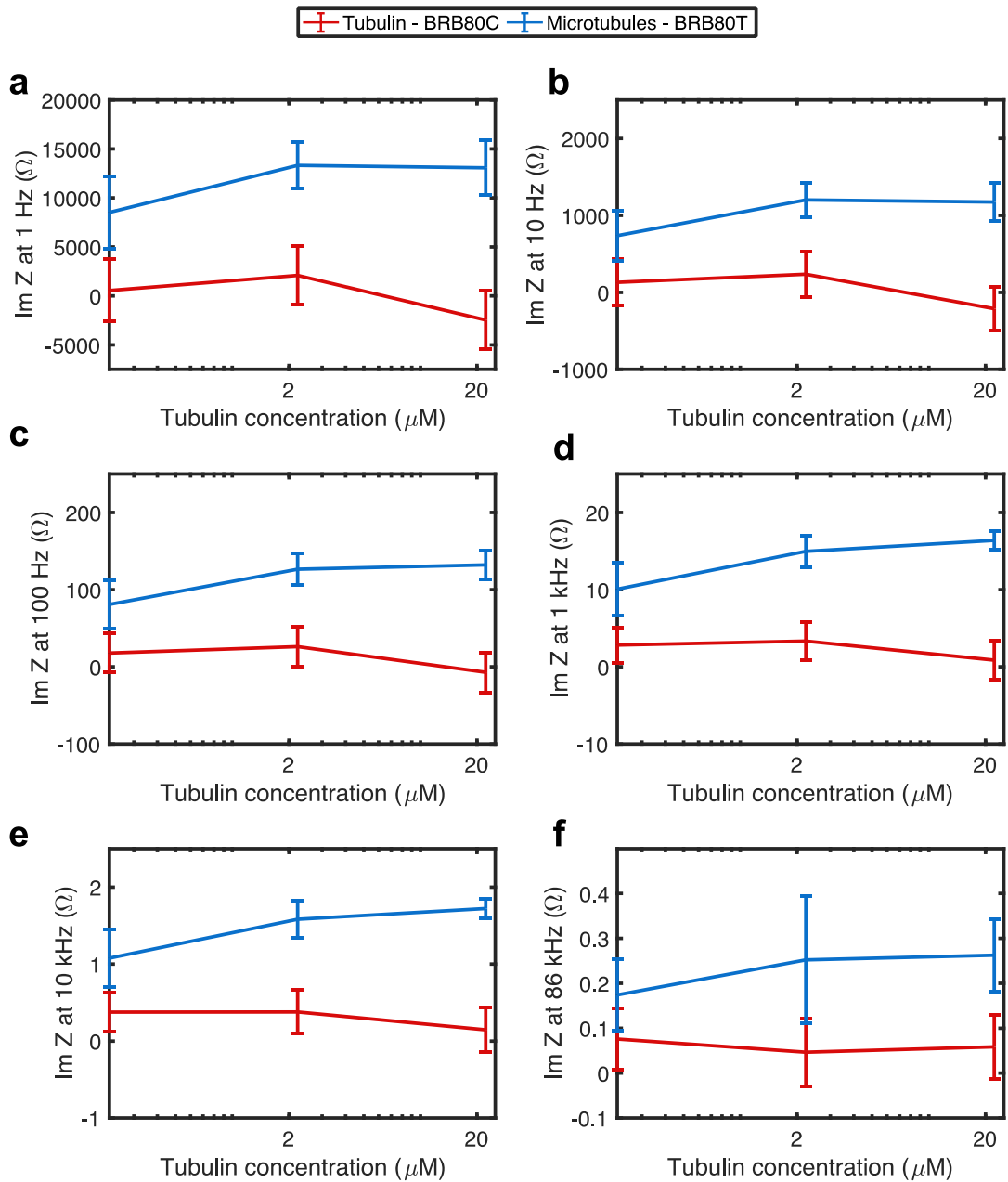

Figure S4. Microtubules increase solution capacitance while unpolymerized tubulin does not. Graphs showing differences in the imaginary component of impedance as a function of tubulin concentration at input AC frequencies of (a) 1Hz, (b) 10 Hz, (c) 100 Hz, (d) 1 kHz, (e) 10 kHz and (f) 86 kHz.

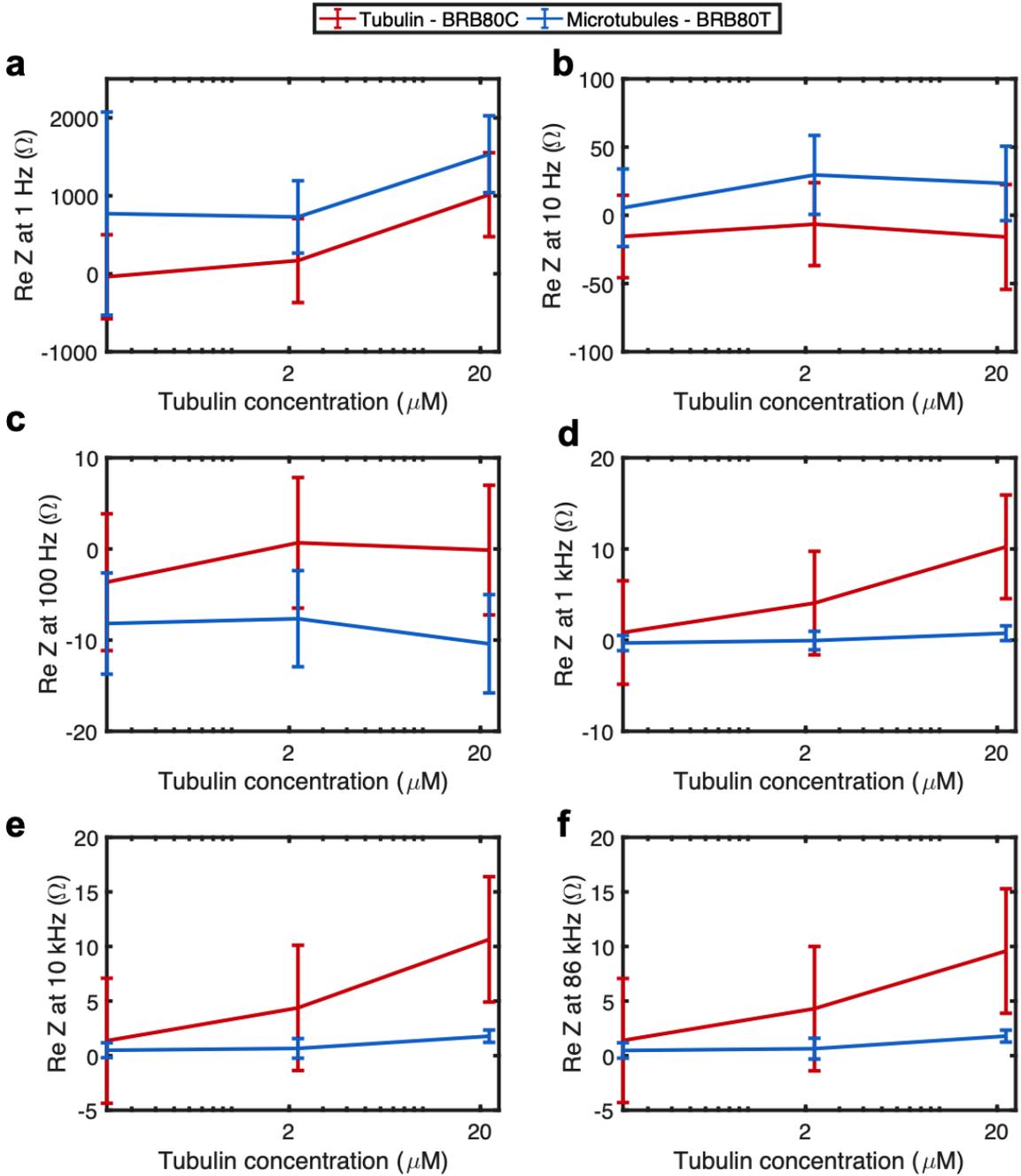

Figure S5. A 'reversal' in the resistive behavior of microtubules is observed between 10 and 100 Hz. Graphs showing differences in the real component of impedance as a function of tubulin concentration at input AC frequencies of (a) 1Hz, (b) 10 Hz, (c) 100 Hz, (d) 1 kHz, (e) 10 kHz and (f) 86 kHz.

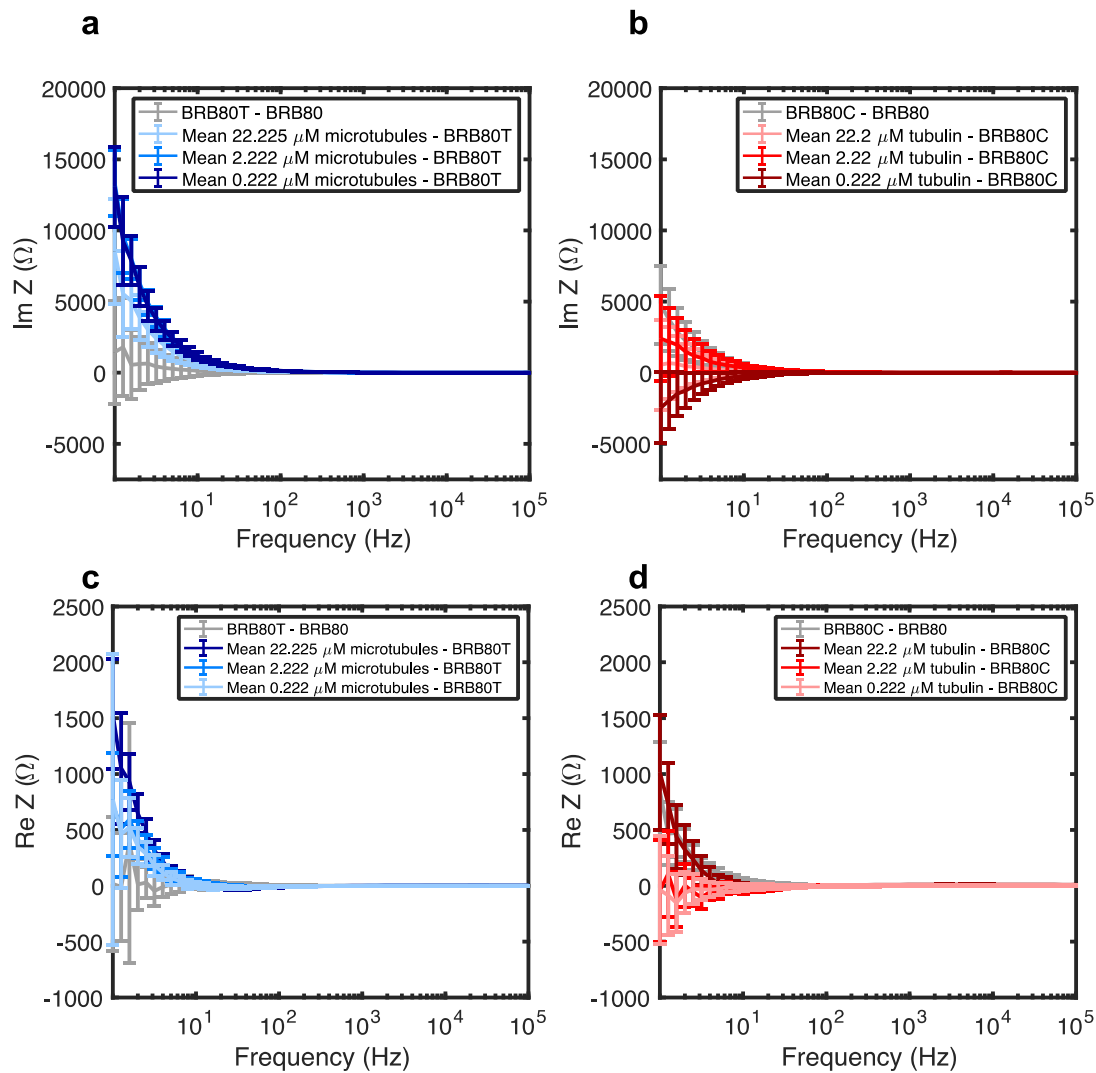

Figure S6. Example of microtubule and tubulin subtraction with backgrounds, to display typical impedance differences. Raw values for (a) imaginary and (b) real component of impedance at different concentrations of microtubules. Raw values for (c) imaginary and (d) real component of impedance at different concentrations of tubulin.

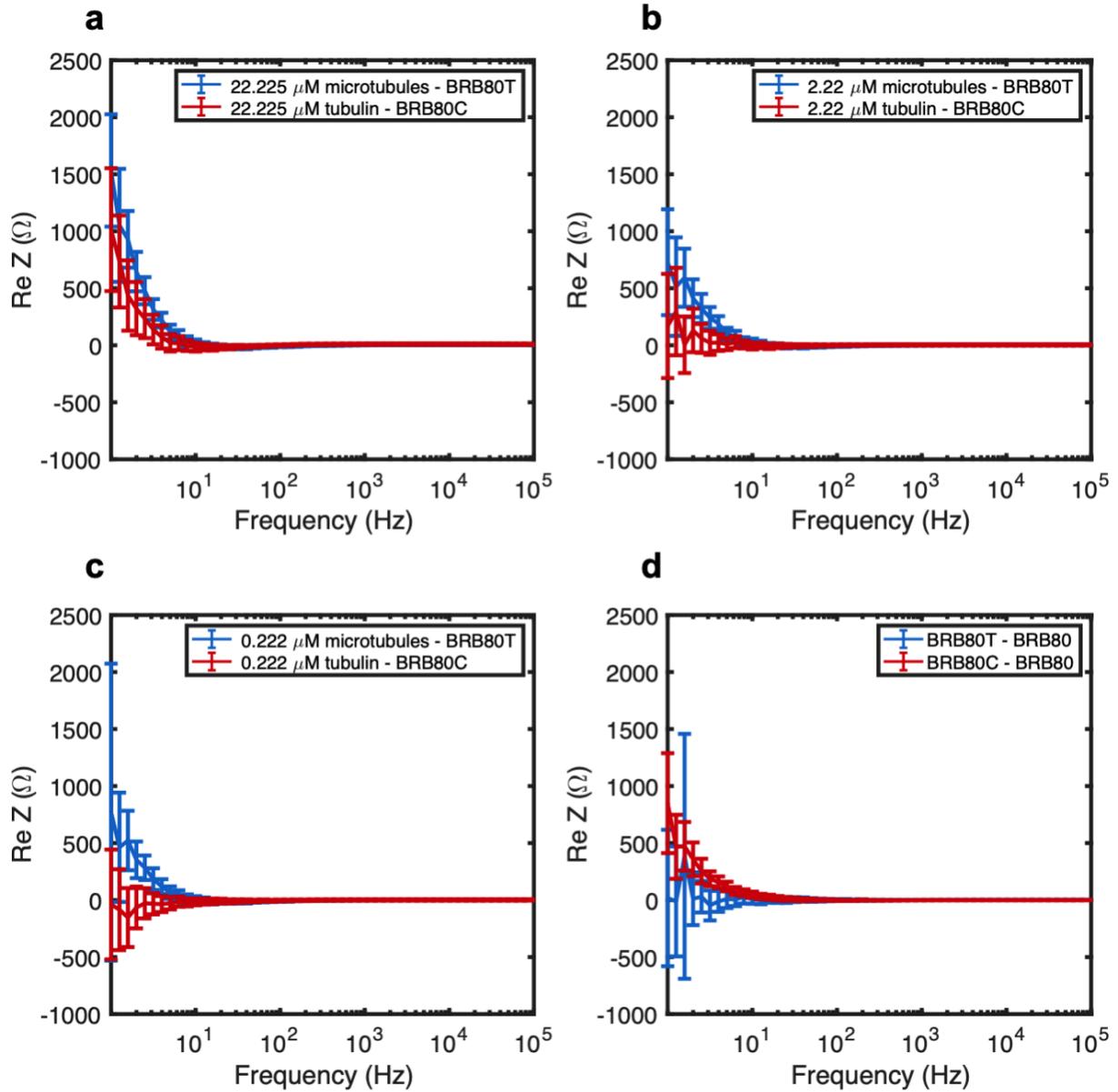

Figure S7. The trend for both microtubules and tubulin displays increasing solution resistance. Graphs showing differences in the real component of impedance as a function of decreasing input AC frequency at total tubulin concentrations of (a) 22.225 µM (n = 22 experiments for tubulin, n = 21 for microtubules), (b) 2.222 µM (n = 35 experiments for tubulin, n = 49 for microtubules) (c) 0.222 µM (n = 35 experiments for tubulin, n = 49 for microtubules), (d) comparison of the effects of paclitaxel (BRB80T) and colchicine (BRB80C, n = 49 experiments for BRB80T, n = 35 for

BRB80C, n = 84 experiments for BRB80). Graphs display average values. Error-bars represent standard deviation.

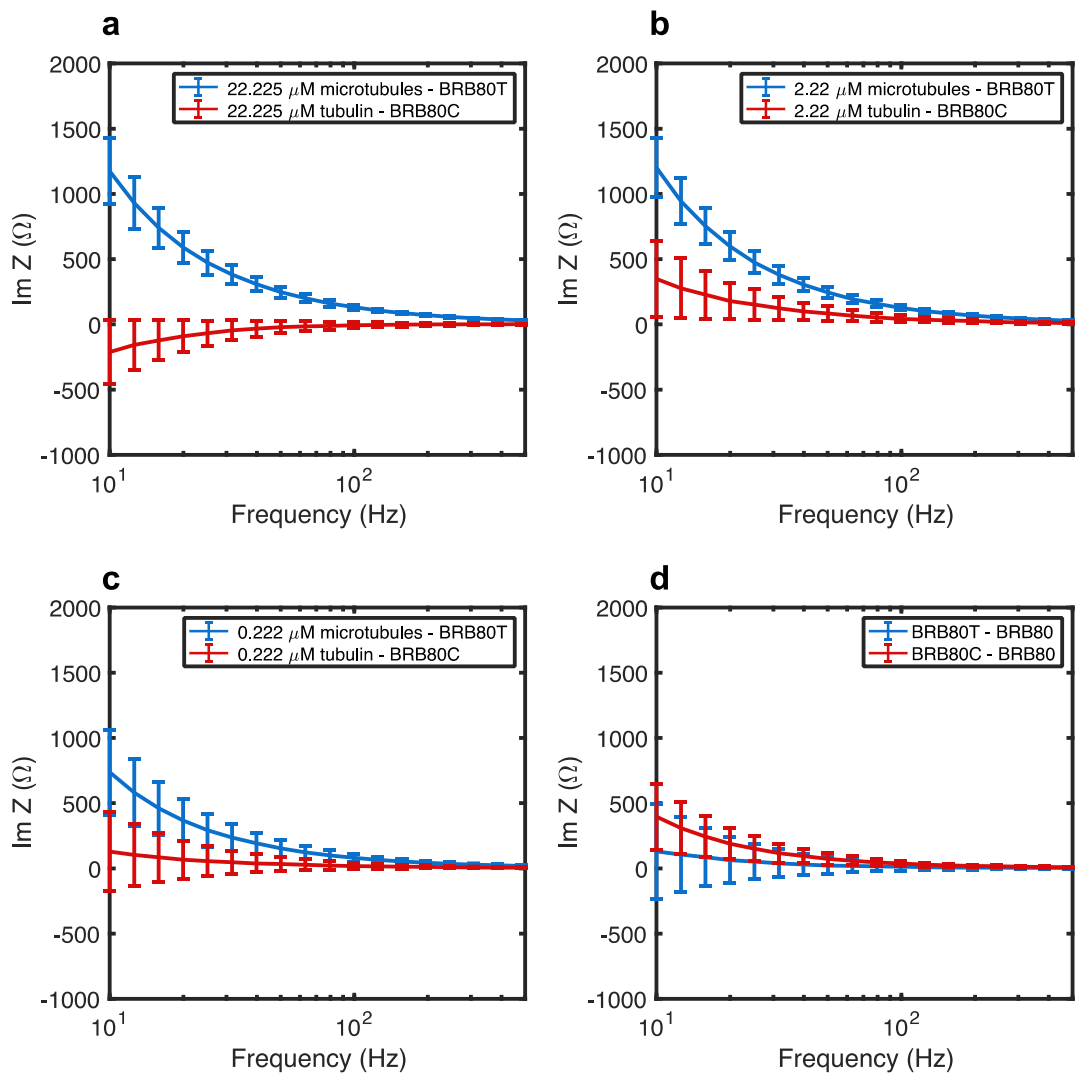

Figure S8. No 'reversal' in the resistive behavior of microtubules is observed between 10 and 100 Hz. Graphs showing differences in the real component of impedance as a function of decreasing input AC frequency at total tubulin concentrations of (a) 22.225 µM, (b) 2.222 µM, (c) 0.222 µM, (d) comparison of the effect of paclitaxel and colchicine on impedance.